# A Perspective on Topological Charge Density Wave Materials

Chun-Tung Lien[1,2], Nidhish Thiruthukkal Puthenveettil[1,2], Alexander A. Balandin[1,2,3]†, Md Shafayat Hossain[1,2,3]†

[1]Department of Materials Science and Engineering, University of California, Los Angeles, California 90095 USA

[2]California NanoSystems Institute, University of California, Los Angeles, California 90095 USA

[3]Center for Quantum Science and Engineering, University of California, Los Angeles, California 90095 USA

†Corresponding author E-mail: balandin@seas.ucla.edu ; shossain@seas.ucla.edu

**Abstract**

Charge density waves are among the earliest known ordered states in quantum materials, yet they have re-emerged at the center of modern discussions of topology, symmetry breaking, and collective dynamics. Their description has expanded far beyond the classical picture of a Fermi-surface instability accompanied by nonlinear transport. In several material classes, charge-density-wave order reshapes electronic bands, boundary states, defect structures, and electromagnetic responses in ways that invite a topological interpretation. Real-space imaging has shown that incommensurate order can form networks of phase defects rather than uniform modulations. Boundary-resolved spectroscopy has revealed in-gap modes whose existence and spatial structure track the charge-density-wave order, although such observations remain confined to a single material family. In Weyl materials, density-wave gaps have been linked to axion-like magnetoelectric response, placing the charge-density-wave phase within a dynamical-field description of the effective action. In kagome metals, charge order is intertwined with Berry curvature, unconventional symmetry breaking, and superconductivity, further connecting charge-density-wave physics with concepts commonly associated with topology. These developments have led to the term "topological charge density wave," but the phrase is used to describe several physically distinct phenomena. We distinguish three meanings: **(i)** topology of the charge-density-wave order parameter and its defects, **(ii)** topology of the charge-density-wave-reconstructed electronic structure and its associated boundary modes, and **(iii)** axion-like electromagnetic response generated by density-wave order. The experimental and theoretical support for these three meanings differs substantially. This Perspective traces the development of these ideas, clarifies the physical meaning and evidence behind each class, identifies unresolved questions and points of debate, and outlines experimental criteria for establishing topology in charge-density-wave states. We conclude by highlighting opportunities to discover, classify, and ultimately control topological charge-density-wave phenomena in quantum materials.

## 1. Introduction

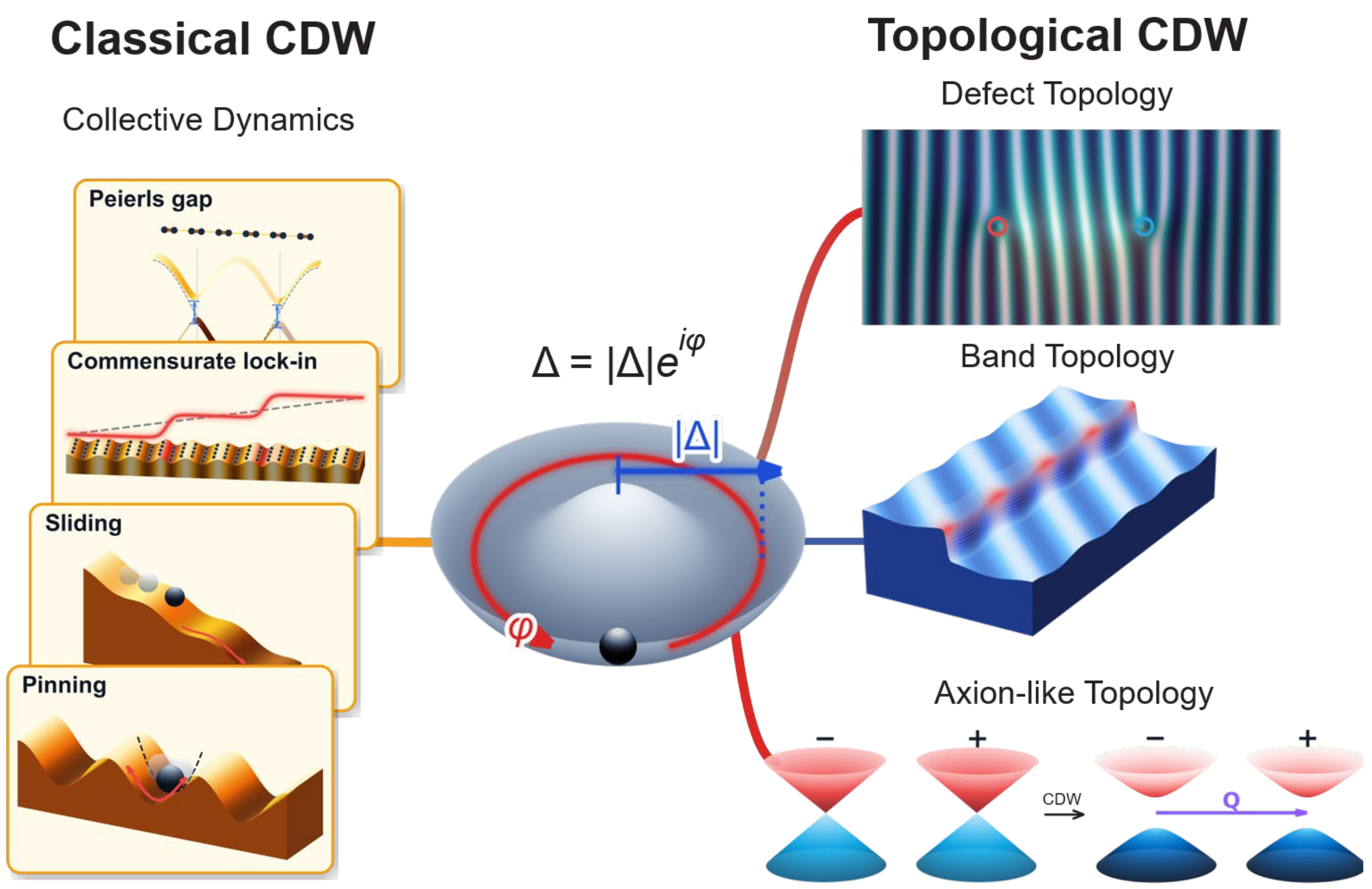

**Figure 1.** An illustration of the evolution of the concept of charge-ordered states in quantum materials. It begins with classical CDW order and collective dynamics. From there, it branches into three contemporary topological interpretations: order-parameter defect topology, band topology and boundary modes, and the potential axionic response in Weyl systems.

Charge density waves (CDWs) are characterized by a periodic modulation of the charge density, breaking translational symmetry and giving rise to a superlattice periodicity directly observable in reciprocal space. Fermi liquids are particularly susceptible to CDW transitions, especially within low-dimensional systems. In these systems, the interplay between electrons and holes at the Fermi energy can lead to Fermi surface nesting, thereby facilitating a CDW transition [1–7]. This transition is accompanied by the opening of a spectral gap at both the Fermi energy and the nesting wavevector, thereby lowering the energy of electrons near the Fermi level.

Notably, the occurrence of a CDW transition is frequently associated with a structural distortion, often via electron-phonon coupling, which stabilizes the nesting conditions[8–13]. For such transitions to occur, the increase in elastic energy must be adequately compensated by a larger reduction in the electronic energy of the system. This principle was elucidated by Peierls, who provided a theoretical framework for the formation of CDWs in one-dimensional (1D) atomic chains, which leads to a metal-insulator transition [14]. Quasi-1D systems are particularly susceptible to these Fermi surface instabilities [15–18].

In principle, the mechanisms governing CDWs are not expected to extend to wavevectors connecting states far from the Fermi energy, as the associated electronic interactions are generally deemed insufficient to induce a phase transition. Thus, CDW distortions are expected to occur predominantly at or near the Fermi energy, reflecting a direct association with Fermi-surface instabilities.

Beyond Peierls' original framework, recent studies have proposed a strong and wavevector-dependent electron-phonon coupling as a driving force for CDW transitions in quasi-two-dimensional (quasi-2D) materials [8,13,19,20]. In such CDW materials, spectral reconstructions may extend across the entirety of the electronic structure, indicating that the CDW transition could represent a metal-metal Lifshitz transition at the Fermi surface [21–27].

Here, a CDW state opens a gap at its corresponding wavevector, thereby reconstructing the Fermi surface and defining a new Brillouin zone. When the Fermi surface is only partially gapped, zone folding within the Brillouin zone delineated by the new CDW periodicity leads to a significantly reconstructed remnant Fermi surface. Hence, prevailing theoretical frameworks associate CDW formation with a Fermi surface, and such transitions typically induce either a metal-insulator or a metal-metal transition at this surface. Interaction-driven and excitonic density-wave states, discussed in Section 4, are important exceptions to this expectation.

The CDW state is typically characterized by a complex order parameter, which determines both the gap scale and the lattice distortion amplitude. The phase of this order parameter acts as a collective coordinate that impurities and boundaries can influence, and it can respond nonlinearly to electric fields. The dynamic implications of this phase, including phenomena such as pinning, depinning, and collective transport, have been organized into quantitative experimental criteria for CDW dynamics [11]. Conversely, the phase field accommodates quantized defects. Near the commensurate–incommensurate transitions, the lowest-energy pathway to incommensuration may occur through localized phase slips rather than a uniform adjustment of the wavevector.

The concept of discommensuration was introduced [28] to describe such a defect, demonstrating that a finite density of phase slips can represent the incommensurate state near lock-in. This defect-based description serves as the minimal framework for many contemporary real-space datasets, where the CDW is more accurately depicted as a textured phase field characterized by domain networks and defect vertices. The study of $NbSe_2$, which interprets incommensuration in terms of a network of topological excitations, exemplifies this regime [29–34].

Topological CDWs have been suggested to emerge when CDW order interacts with electronic structures that already exhibit topological band characteristics. Several classes of topological materials host nontrivial band topologies that can couple to CDW order. Representative examples include the Type-III Weyl semimetal $Ta_2Se_8I$ [35,36], the $Z_2$ topological kagome metals $AV_3Sb_5$ (A = K, Rb, Cs) [37–43] with pronounced Berry curvature near van Hove singularities, and enforced semimetals with Fermi-point degeneracies (ESFD) such as the transition-metal dichalcogenides (TMDs) 2H-$NbSe_2$ [29,44]. Quasi-one-dimensional transition-metal trichalcogenides (TMTs) further exhibit Dirac- or Weyl-like band features and, in selected members, signatures consistent with topological superconductivity [45,46]. In addition, moiré heterostructures based on TMDs realize tunable nonzero Chern numbers and fractional Chern insulator states at zero magnetic field [47–49].

These characteristics encompass non-trivial electronic band structures, such as protected boundary states, Berry curvature, non-zero Chern numbers, and anomalous electromagnetic responses, which can be reshaped or coupled with CDWs. In this context, a CDW order does more than merely fold bands and open gaps; it can reorganize boundary spectra. This reorganization occurs at the boundary, where the effects of topology are tested and where the CDW phase is influenced by termination, pinning, and local symmetry.

A commonly studied experimental platform for these phenomena is $Ta_2Se_8I$. Here, scanning tunneling microscopy has revealed in-gap spectral weight localized at step edges within the CDW state, with spatial modulation linked to the CDW wavevector [35]. This finding establishes a direct connection between charge order and boundary electronic structures in a topological material [35,36,50–52]. Several studies on $Ta_2Se_8I$ have associated CDWs with topological features through electromagnetic responses rather than through boundary spectra [51].

Some reports have documented magnetotransport signatures in the sliding CDW regime that resemble axionic CDW responses, where the CDW phase mode is treated as a dynamical field in the effective action. Other studies have interpreted similar phenomena without incorporating a topological framework [53]. These differing perspectives have led to controversies that necessitate carefully defined experimental criteria to gather further evidence and ascertain whether topological non-triviality truly exists.

Kagome metals present a unique environment where charge order, band geometry, and superconductivity converge. In $KV_3Sb_5$, scanning tunneling microscopy has revealed an unconventional chiral charge order characterized by a robust 2×2 superlattice and a spectroscopic gap structure associated with this modulation [37,40,54]. This finding lays a foundation for discussing symmetry breaking beyond a simplistic scalar-density modulation. The broader $AV_3Sb_5$ family has spurred intense research activity, as the kagome band structure positions the Fermi level near van Hove singularities, leading to significant Berry curvature that coexists with charge order and superconductivity [39,43,55–59].

Notable advances have also been made in the study of CDW transport under external electric fields or currents. In 1T-$TaS_2$, the development of integrated heterostructures has enabled reproducible room-temperature threshold behavior and oscillatory dynamics in compact circuits, facilitating controlled access to the transition from nearly commensurate to incommensurate states [60–63]. Subsequent work employed high-frequency oscillation measurements and nanosecond pulsing to discern intrinsic switching pathways from self-heating, thereby connecting oscillatory behavior to established CDW dynamical concepts [64–67]. Additionally, noise spectroscopy has introduced a sensitive probe for assessing depinning and phase kinetics in thin films, particularly relevant when transport is used to support topological characteristics [68–71].

To facilitate our discussion, we organize the manuscript in the following manner. Section 2 establishes the classical understanding of the CDW as a condensate with collective dynamics. Sections 3 to 5 focus on three distinct classes of topological CDW, organized by us to show how topology is relevant in CDW systems (Fig. 1). Section 3 focuses on the topology of the CDW phase field in real space, examining discommensurations and the vertices of domain networks (Class I). Section 4 discusses the topology of the reconstructed electronic structure, especially at boundaries and defects, using $Ta_2Se_8I$ as a notable case study. This section concludes by exploring interaction-driven charge order in topologically nontrivial flat bands (Class II). Section 5 addresses topology in low-energy responses, specifically axion-like magnetoelectric transport in Weyl systems, and outlines the necessary constraints to exclude non-topological alternatives (Class III).

Table 1 delineates this classification. Section 6 examines nonequilibrium control of these three classes of topological CDW and the corresponding devices. Section 7 revisits ongoing discussions where terminology and evidence often diverge, and the corresponding open questions in the field, and Section 8 outlines measurement and theoretical targets that could improve predictive capabilities for topological CDWs.

| **Class** | **Where the topology resides** | **Order parameter / invariant** | **Level of evidence** | **Representative materials** |
|---|---|---|---|---|
| **I**<br>**Order-parameter topology** | In the CDW phase field itself. The ordered state hosts defects across which the phase advances by a quantized amount. The electronic structures do not need to be topological. | $\Delta(r) = \lvert\Delta\rvert e^{(i\varphi(r))}$; winding of φ. $\pi_1(S^1) = \mathbb{Z}$ for single-Q; $\mathbb{Z}_3$ vertices for triple-Q. Defect density sets $Q - Q_C$ (McMillan relation). | **Established.**<br>Phase field imaged directly; quantized phase shifts resolved; defect density tracks the diffraction wavevector. | 2H-$NbSe_2$[31,34,44,72]; $LaTe_3$[73,74]; In/Si(111)[75]; $UTe_2$[76,77] |
| **II**<br>**CDW-reconstructed band topology and boundary modes** | In the single-particle spectrum of the ordered state. Zone folding and gapping at ±Q change band connectivity, producing a nontrivial invariant and/or boundary states present only below $T_{CDW}$. | $\mathbb{Z}_2$ indices, Chern and mirror Chern numbers, band-inversion parity eigenvalues and symmetry indicators are all evaluated in the folded (CDW) zone.<br>Bulk–boundary correspondence gives the edge spectrum; boundary-state period locked to $\lambda_{CDW}$. | **Inconclusive; material-specific.**<br>Reconstruction established. CDW-locked boundary modes reported in specific materials but lacks evidence on others. | $Ta_2Se_8I$[35,78]; $Ta_2NiSe_7$[79,80]; $AV_3Sb_5$[25,38–42,81]; FeGe[82–84]; $VTe_2$[85]; $TaTe_2$[86]; monolayer 1T-$TiSe_2$ [87–89]; moiré superlattices [47–49]. |

| Class | Where the topology resides | Order parameter / invariant | Level of evidence | Representative materials |
|---|---|---|---|---|
| **III** **Axion-like (θ-term) response** | In the low-energy electromagnetic response rather than the spectrum. If Q connects Weyl nodes of opposite chirality, the sliding phase becomes a dynamical axion field entering the effective action through a θ E·B term. | $\theta(r, t) = 2Q\cdot r - 2\varphi_{CDW}$, the CDW phase promoted to a dynamical field. Chirality $\chi = \pm 1$ of the connected nodes; chiral anomaly coefficient. | **Contested.** Classical sliding-CDW dynamics reproduce several of the signatures; node-connecting nesting is not confirmed by ARPES. | $Ta_2Se_8I$[35,51,90–93] |

**Table 1. A classification on different meanings of topological charge density wave.**

## 2. Classical foundations: CDWs as condensates with collective dynamics

The early phenomenology surrounding CDWs was driven by observations in layered transition-metal dichalcogenides, where anomalies in transport and diffraction were unified under the concept of a periodic lattice distortion correlated with a modulation of electronic density. Pioneering studies played a crucial role in establishing this coherent framework for the dichalcogenides [94–97]. A key insight from these works is that a CDW can be characterized by a complex order parameter comprising both amplitude and phase. In this context, the phase represents a low-energy field responsible for sliding, pinning, and defect formation.

The dynamical theory focuses on concepts such as pinning, depinning, nonlinear transport, and the frequency-dependent response of a pinned condensate [11]. While these classical concepts remain relevant today, advances in understanding topological materials and their transport properties [56,57,98–100] have introduced alternative interpretations of similar phenomena. For example, negative magnetoresistance and anomalous Hall effects arise from topological features like Berry curvature or chiral anomalies, making it challenging to separate them from purely phonon-driven nesting effects. To distinguish between classical phenomena associated with band structure and topological impacts, it is essential to consider domain dynamics and nonequilibrium pathways as integral components of the underlying physics.

## 3. Topology enters through commensurability: discommensurations and phase defects

Interestingly, the first rigorous appearance of topology in CDW physics did not emerge from the typical discussion surrounding band invariants, which usually serves as the primary approach to examining topological band structures. Instead, early research derived topology from the structure of the CDW phase field near commensurate–incommensurate transitions [28,101,102]. A CDW is characterized by a complex order parameter $\Delta(r) = |\Delta| e^{i\phi(r)}$. When the ordering wavevector aligns commensurately with the lattice, the phase $\phi$ is constrained to specific values determined by the periodicity of the lattice. Conversely, in the incommensurate regime, the phase is free to slide relative to the lattice.

The transition between these two states cannot always be described by a uniform shift in the modulation wavevector. It requires a quantitative framework of the theory of discommensurations that was developed for layered dichalcogenides [28]. It was demonstrated that near a lock-in transition, which represents a phase change from incommensurate to commensurate, the energetically favorable configuration may consist of locally commensurate regions separated by narrow phase-slip walls—these discommensurations allow the CDW phase to slide by $2\pi/3$ to accommodate incommensuration. Across each wall, the CDW phase shifts by a fixed fraction of $2\pi$. Consequently, the incommensurate state can be constructed from a periodic arrangement of these phase slips. The density of discommensurations dictates the extent to which the average wavevector deviates from the commensurate value.

This finding established that incommensurability can be represented by a finite density of topologically stable phase defects, rather than through a uniform modulation. In the language of the earlier literature, a discommensuration is a phase soliton, stabilized by the commensurability potential rather than by any property of the electronic spectrum. The order parameter can take values on a circle, which means that the relevant homotopy classification for a single-Q modulation is $\pi_1(S^1) = \mathbb{Z}$. The integer that labels a wall indicates how many times the phase winds across it. In two dimensions, triple-Q order extends this classification: the points where three domain walls meet are referred to as $\mathbb{Z}_3$ point defects. Recent real-space experiments have revealed that it is the network of these vertices, rather than the isolated walls, that is significant (Fig. 2).

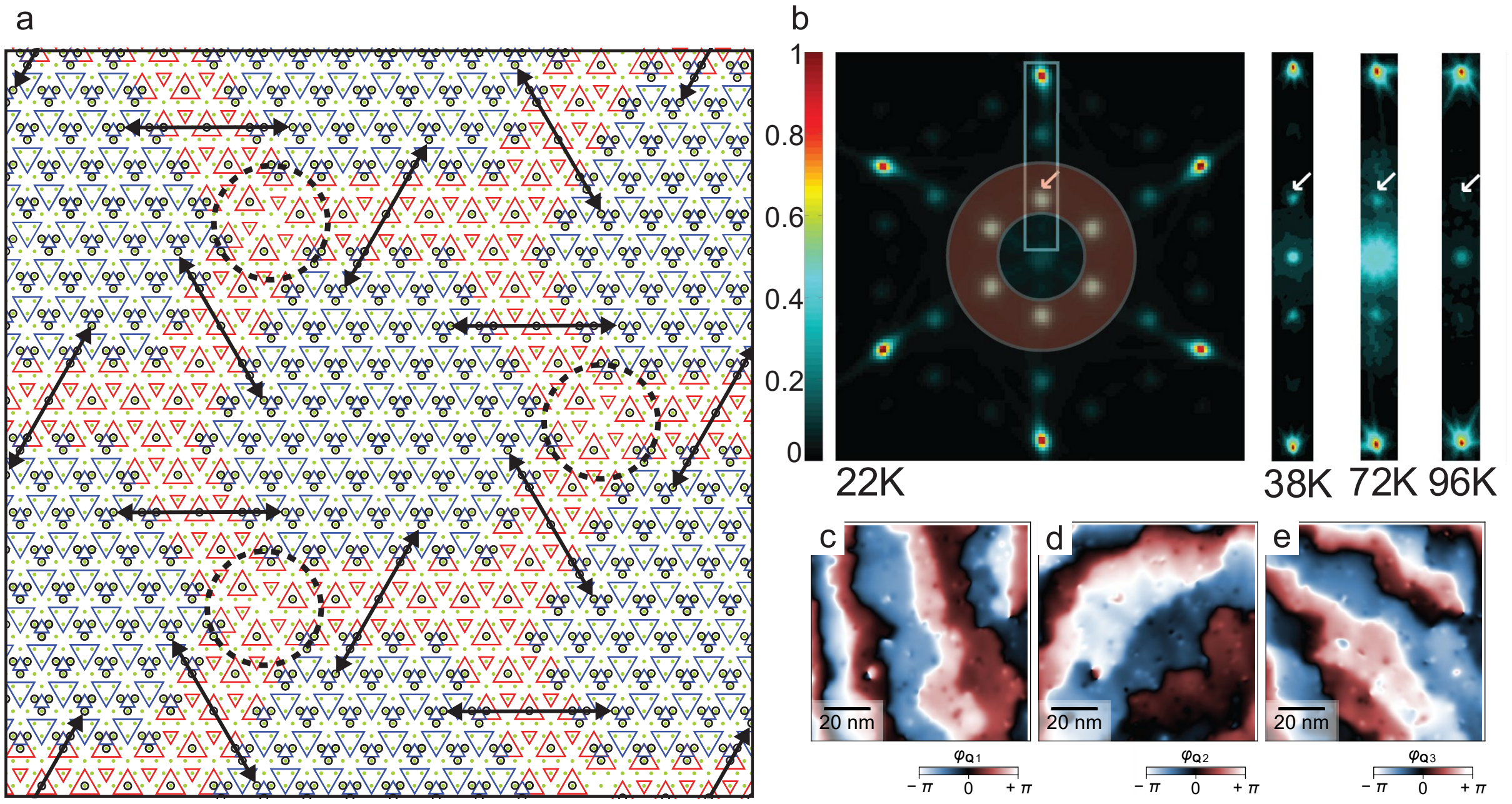

**Figure 2. A visualization of the real-space topology effect in 2H-NbSe$_2$. (a)** Schematic illustration of the ideal 2H-NbSe$_2$ honeycomb lattice. The black dashed circle denotes the surface

topological vertices, and the arrows indicate the direction of CDW translation. Reprinted figure with permission from G. Gye, E. Oh, and H. W. Yeom, Phys. Rev. Lett. 122, 016403 (2019), https://doi.org/10.1103/PhysRevLett.122.016403. Copyright (2019) by the American Physical Society. [34] **(b)** Fourier transform in reciprocal space showing the CDW wave vectors. Red peaks represent the atomic Bragg peaks, and the peaks indicated by white arrows correspond to the CDW wave vectors. Reprinted figure with permission from C. J. Arguello, S. P. Chockalingam, E. P. Rosenthal, L. Zhao, C. Gutiérrez, J. H. Kang, W. C. Chung, R. M. Fernandes, S. Jia, A. J. Millis, R. J. Cava, and A. N. Pasupathy, Phys. Rev. B 89, 235115 (2014), https://doi.org/10.1103/PhysRevB.89.235115. Copyright (2014) by the American Physical Society. [103] **(c)-(e)** Visualization of topological defects (soliton pairs) and the induced CDW domain walls with different phase shifts φ. Reprinted figure with permission from S. Yoshizawa, K. Sagisaka, and H. Sakata, Phys. Rev. Lett. 132, 056401 (2024), https://doi.org/10.1103/PhysRevLett.132.056401. Copyright (2024) by the American Physical Society. [33]

This framework presents two key features. First, a discommensuration is stable because the phase change across it is quantized relative to a commensurate reference. Second, the energy cost associated with defects is governed by the competition between phase stiffness and the lattice lock-in potential. These elements are sufficient for classifying defects and for computing their spacing as a function of temperature or control parameters. Modern real-space probes have validated that this defect-based description is not confined to idealized models.

In a one-dimensional system, measurement focuses on the individual defect rather than the network itself. In indium atomic wires on a Si(111) substrate, the ground state of a coupled double Peierls chain is fourfold degenerate. The solitons that connect these states exhibit chirality, which allows them to be imaged and switched one at a time [75]. Scanning tunneling microscopy studies of 2H-$NbSe_2$ have revealed that the incommensurate CDW can be organized into networks of phase defects and domain structures rather than by a single, spatially uniform wavevector [30,33,34]. In particular, competing CDW configurations can coexist, forming a textured phase landscape whose vertices act as topological excitations that mediate the effective incommensuration. This interpretation directly links diffraction wavevectors to defect density and domain arrangement.

While diffraction provides an average ordering wavevector, real-space imaging resolves the underlying phase field. When the two are combined, incommensurability is understood as a measurable consequence of defect topology. The same description of phase defects applies to the rare-earth tritellurides. In $LaTe_3$, ultrafast photoexcitation disrupts and subsequently re-establishes the modulation by generating topological defects in the order parameter.

In this context, the density of these defects serves as an observable that tracks the transition [73,74]. However, the two probing methods do not always align. Scanning tunneling microscopy on $UTe_2$ reveals a charge modulation that dissolves as defect pairs with opposite winding form, a phase-field signature as previously described [76]. In contrast, bulk resonant elastic X-ray scattering does not detect any charge order in the normal state of the same compound [77]. This disagreement is still unresolved, highlighting the importance of reporting surface-resolved phase imaging alongside bulk diffraction results. Additionally, in the rare-earth tritellurides $RTe_3$, detecting an axial Higgs mode indicates that collective modes of the CDW can exhibit nontrivial symmetry characteristics [104].

This framework requires no reference to band topology or Berry curvature, relying solely on the structure of the CDW phase and the energetics of commensurate locking. This defect-based topology remains central to contemporary discussions of topological CDWs [31,35,73,74,105]. If a CDW state carries nontrivial boundary modes or associated transport or optical responses, the phase texture of that ordered state must also be examined [35,50,100,106]. Discommensurations, domain walls, and defect vertices modify the local gap structure and can host bound states independent of global band invariants.

For this reason, the commensurate–incommensurate problem provides the historical and conceptual foundation for all later uses of topology in CDW systems. It is important to clarify what this distinction does not imply. Vortices, domain walls, and solitons can be found in many conventional broken-symmetry phases without those phases being classified as topological states of matter [107–110]. The same principle applies here: a CDW that contains topological defects is not automatically classified as a topological CDW. A defect-bearing order parameter differs significantly from an ordered state characterized by nontrivial electronic topology.

## 4. Band topology reshaped by CDW order: from reconstruction to boundary modes

A CDW reconstructs the electronic structure by enlarging the unit cell and folding the Brillouin zone. In conventional scenarios, this reconstruction typically results in a partial gapping of the Fermi surface along with a redistribution of spectral weight [72,111–113]. However, in materials exhibiting nontrivial band topology, the implications of the same reconstruction can be qualitatively different. The CDW order parameter couples electronic states that are separated by the ordering wavevector Q. When these states arise from bands with nontrivial topology, the resulting density-wave gap can connect, split, or invert band features in such a way that alters the boundary spectra [35,85,88].

Band reconstruction and topological reconstruction must be distinguished. A purely structural reconstruction modifies dispersion relations but does not necessarily induce boundary states. In contrast, a topological reconstruction occurs when the ordered phase supports boundary or defect-bound modes that are linked to the symmetry and band structure of the CDW state. Establishing this connection necessitates boundary-sensitive spectroscopy with both momentum and spatial resolution.

$Ta_2Se_8I$ is a compelling example. This quasi-one-dimensional compound features a CDW ground state and showcases topological band characteristics in its normal phase. Scanning tunneling microscopy and spectroscopy have identified an energy gap associated with CDW order in the bulk, as well as, notably, in-gap spectral weight localized at step edges (Fig. 3) [35]. In contrast, tunneling spectroscopy performed at CDW edge dislocations in the same compound demonstrates the bulk gap without any in-gap modes [92], indicating that the boundary signal is specific to the step edges. The boundary state tunneling conductance (dI/dV) shows modulation consistent with a wavevector equal to the CDW ordering wavevector in real space, and its energy position evolves in accordance with the bulk CDW gap across varying temperatures. These measurements illustrate that the boundary electronic structure is reorganized in the CDW phase, with this reorganization being spatially locked to the charge order. Consequently, the data provide a concrete example of how the CDW order and boundary modes are directly linked in a material with topological band features.

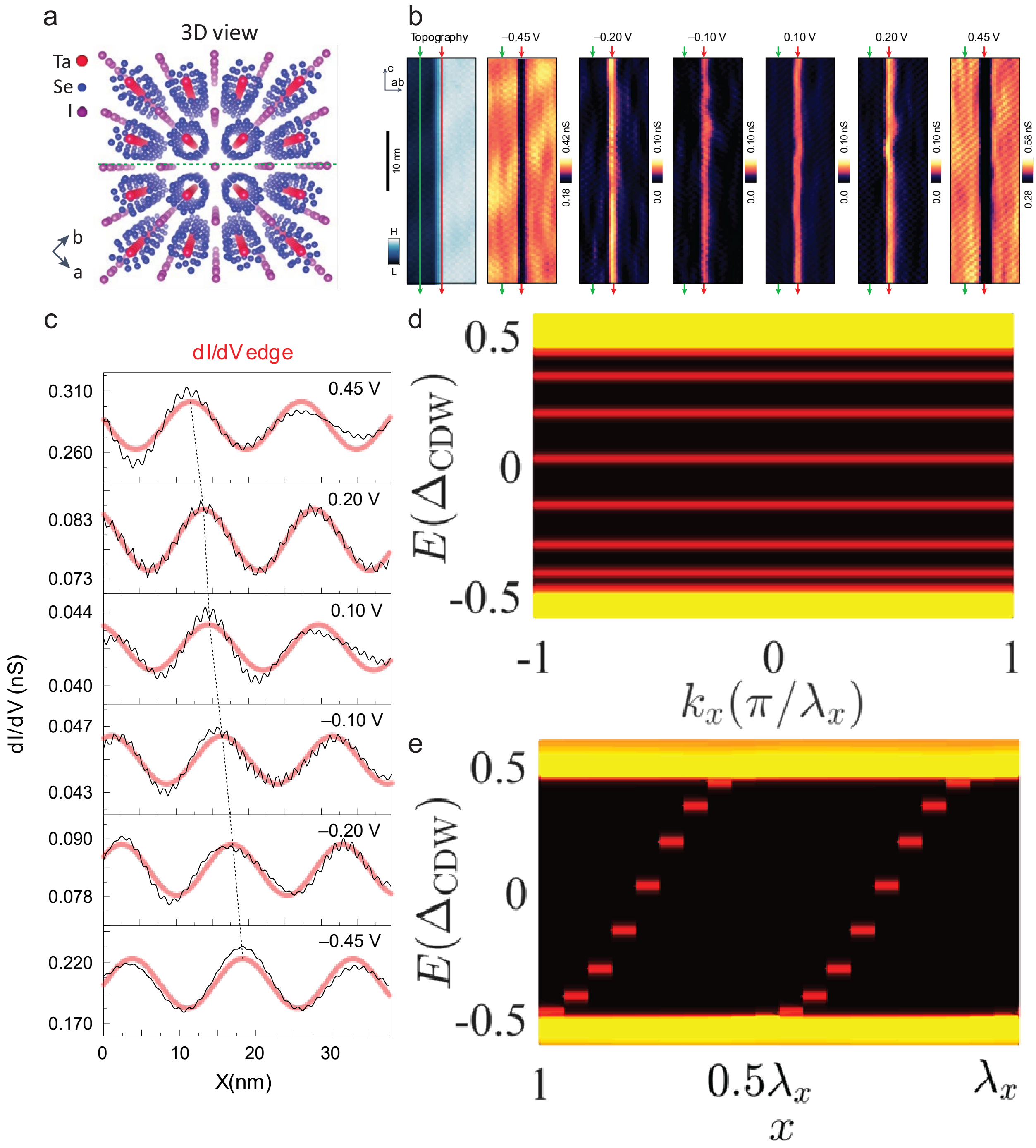

**Figure 3. CDW-locked edge mode on $Ta_2Se_8I$. (a)** $Ta_2Se_8I$ crystal structure, the probing is conducted on the natural cleaving plane labeled by green dashed line. **(b)** A monolayer step edge and its corresponding dI/dV maps, showing a CDW-locked edge state. **(c)** Gradual change of CDW phase along the step edge (marked with a red vertical line in **(b)**), showing that the edge conductance feature disperses in energy as the CDW phase advances along the step edge. Panels **(a)**-**(c)** are reprinted with permission from M. Litskevich et al., "Boundary modes of a charge density wave state in a topological material," Nat. Phys. 20, 1253 (2024), Springer Nature. [35] **(d),**

**(e)** CDW edge modes in momentum space and real space, showing dispersionless behavior in k-space and spectral pseudo-flow along the edge. Reprinted figure with permission from S.-B. Zhang, X. Liu, M. S. Hossain, J.-X. Yin, M. Z. Hasan, and T. Neupert, Phys. Rev. Lett. 130, 106203 (2023), https://doi.org/10.1103/PhysRevLett.130.106203. Copyright (2023) by the American Physical Society. [50]

Two experimental criteria of topological CDW are fulfilled in this case. First, the in-gap boundary spectral weight appears exclusively in the CDW phase and vanishes above the transition temperature. This effectively rules out trivial surface states that are unrelated to charge order. Second, the spatial modulation of the boundary signal corresponds precisely to the CDW periodicity, establishing a connection between the boundary mode and the order parameter rather than attributing it solely to structural termination. These criteria provide a reference point for future experiments investigating CDW-induced topological boundary states.

In sharp contrast, a related quasi-one-dimensional candidate, $Ta_2NiSe_7$, does not currently satisfy either criterion. Angle-resolved photoemission reveals the density-wave gap and backfolded spectral weight below the CDW transition temperature ($T_{CDW} \approx 60$ K), but it detects no surface states. This leaves the expected $\mathbb{Z}_4 = 3$ topological classification for this compound unconfirmed [79,114,115]. Although edge states have been reported and used in saturable absorbers [80], no prior work has shown their confinement to the ordered phase or a spatial period locked to $\lambda_{CDW}$. Thus, $Ta_2NiSe_7$ remains a candidate for this class rather than a confirmed realization.

In a complementary case, $TaTe_2$ shows strong charge modulation and competition between a 3×1 stripe phase and a low-temperature 3×3 modulation [116]. Distinct CDW polymorphs of the same compound are calculated to have different topological characteristics, indicating that the specific ordered configuration determines the band topology [86]. This makes $TaTe_2$ an attractive but untested, target for boundary-resolved spectroscopy.

A related, yet distinct context is presented by kagome metals such as $KV_3Sb_5$ [25,40,81]. In these materials, CDW emerges within a band structure characterized by significant Berry curvature and van Hove singularities near the Fermi level. Scanning tunneling microscopy has revealed a 2×2 superlattice and an unconventional chiral CDW pattern in $KV_3Sb_5$, along with spectroscopic gap features associated with this modulation (Fig. 4). Thus, the CDW reconstructs bands that already possess nontrivial geometric properties. Notably, despite extensive searches, no boundary mode has been detected in this family of materials, in contrast to similar materials such as kagome FeGe [82–84]. Outside the kagome family, a planar Hall response attributed to CDW-induced Berry curvature has been reported in the square-net semimetal $LaAgSb_2$ [117], and planar and nonlinear Hall effects attributed to Berry curvature and its dipole emerge in the commensurate CDW phase of 1T-$TaS_2$ [118].

These developments bring conceptual shifts that deserve emphasis. In materials with nontrivial band topology, CDW order can act as a symmetry-breaking field that alters topological invariants and boundary spectra. Establishing such an alteration in any given material requires combined bulk and boundary measurements that account for temperature dependence and spatial correlation with the CDW modulation. Where these conditions are met, the CDW can reshape the topological character of the electronic structure and leave a distinct imprint on the experimentally accessible boundary modes; where they are not, the evidence supports coexistence rather than causation.

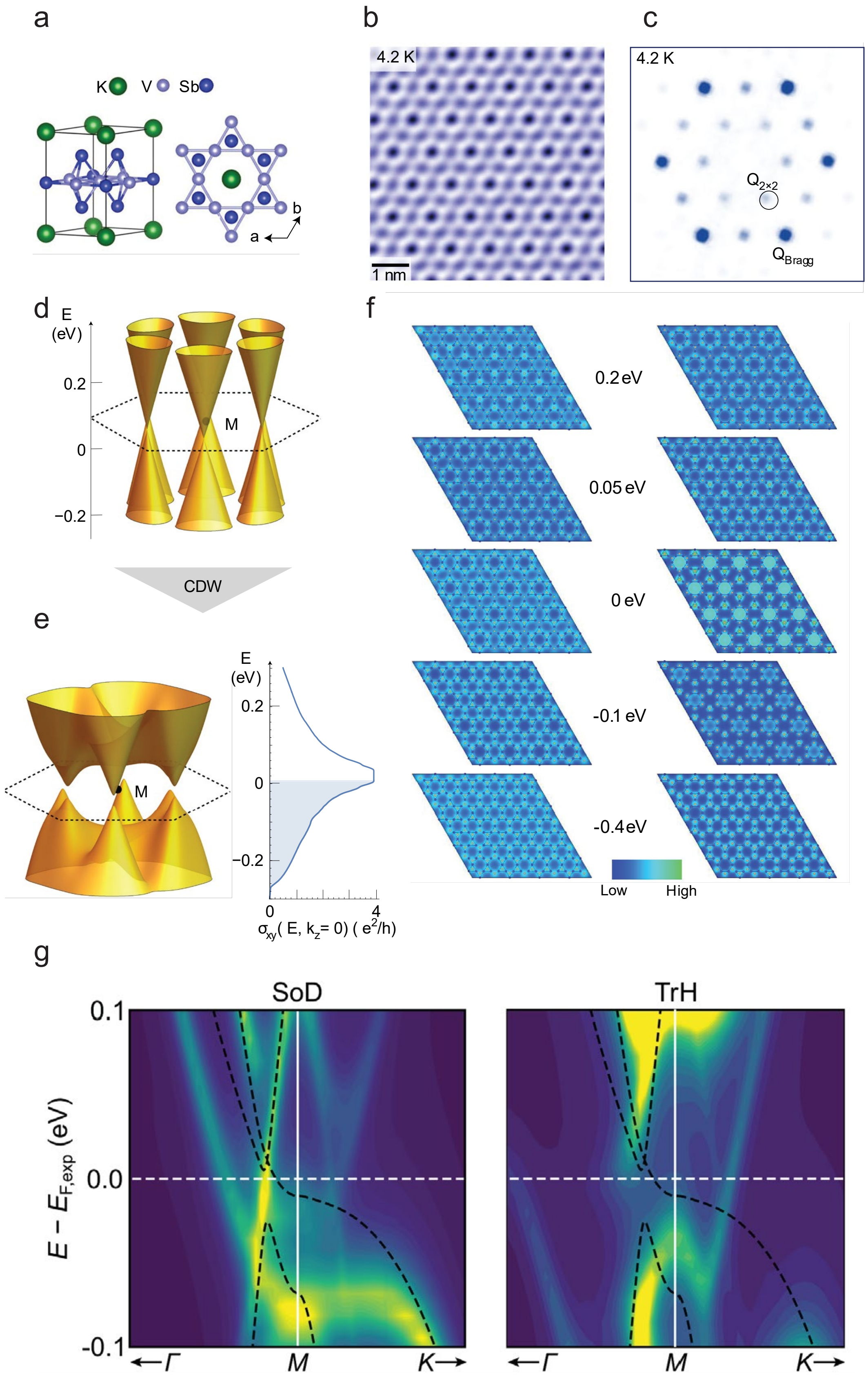
a
K V Sb
b
a
b
4.2 K
1 nm
c
4.2 K
Q2×2
QBragg
d
E (eV)
0.2
0
−0.2
M
CDW
e
M
E (eV)
0.2
0
−0.2
0
4
σxy( E, kz= 0) ( e2/h)
f
0.2 eV
0.05 eV
0 eV
-0.1 eV
-0.4 eV
Low
High
g
SoD
TrH
0.1
0.0
-0.1
E − EF,exp (eV)
←Γ
M
K→

**Figure 4. CDW induced band reconstruction on $KV_3Sb_5$ kagome superlattices. (a)** Three-dimensional view and top view of $KV_3Sb_5$ crystal structure. **(b)(c)** A topographic image of a large Sb surface **(b)** showing a 2 × 2 modulation, and its Fourier transform **(c)** with clear ordering peaks ($Q_{2\times2}$), and Bragg peaks ($Q_{Bragg}$). **(d)** Effective band structure of the Dirac nodal lines formed by the vanadium $d_{xz}/d_{yz}$ orbitals at $k_z = 0$ thanks to the 2 × 2 folding of the Brillouin zone, and **(e)** the topological energy gap at the Dirac cones opened by the unconventional CDW order, with the existence of large Berry curvature. Panels **(a)-(e)** are reprinted with permission from Y.-X. Jiang et al., "Unconventional chiral charge order in kagome superconductor $KV_3Sb_5$," Nat. Mater. 20, 1353 (2021), Springer Nature. [40] **(f)** Simulated STM images for the Star of David and the inverse Star of David CDW phases of $KV_3Sb_5$. Reprinted figure with permission from H. Tan, Y. Liu, Z. Wang, and B. Yan, Phys. Rev. Lett. 127, 046401 (2021), https://doi.org/10.1103/PhysRevLett.127.046401. Copyright (2021) by the American Physical Society. [81] **(g)** Ab initio electronic structure of $CsV_3Sb_5$, illustrating the impact of CDW order and the associated structural distortion on the Dirac-like bands. Reproduced from B. R. Ortiz, S. M. L. Teicher, L. Kautzsch, P. M. Sarte, N. Ratcliff, J. Harter, J. P. C. Ruff, R. Seshadri, and S. D. Wilson, "Fermi Surface Mapping and the Nature of Charge-Density-Wave Order in the Kagome Superconductor $CsV_3Sb_5$," Phys. Rev. X 11, 041030 (2021), https://doi.org/10.1103/PhysRevX.11.041030; licensed under a Creative Commons Attribution (CC BY) license (https://creativecommons.org/licenses/by/4.0/). [25]

Charge order can arise primarily from strong Coulomb interactions in systems with narrow or flat electronic bands, where electron-electron repulsion dominates kinetic energy [119], [120]. Unlike the classical CDWs discussed in Section 2, which depend on electron-phonon coupling and Fermi surface nesting to produce periodic modulations, this type of order forms without a periodic lattice distortion. It can also carry a quantized Hall conductivity along with boundary modes [121]. Such states are sometimes referred to as emergent topological states. We use this term specifically to denote a quantized Hall response that occurs in conjunction with translational symmetry breaking; it does not imply whether topological order is associated with fractionalized excitations.

Early studies of these phenomena required extreme conditions to engineer flat bands. For instance, partially filled Landau levels lead to fractional quantum Hall states [122] [123], and the Hofstadter spectrum generates fractal bands with integer or fractional Chern numbers [124–127]. However, both scenarios necessitate strong magnetic fields and low temperatures, which limit scalability. Moiré superlattices have alleviated this issue. In twisted monolayer-bilayer graphene, charge order at half-integer moiré band filling results in tunable Chern numbers with meV-scale gaps and quantized Hall conductivity at zero magnetic field (Fig. 5) [48]. In this case, charge order generates the invariant rather than modifying an existing one, and the invariant remains a Chern number of the reconstructed band.

Furthermore, related interaction-driven phases with tunable Chern numbers, including fractional states, have been reported in twisted bilayer $MoTe_2$ and related heterostructures [47,49,128,129]. By removing the requirement for a magnetic field, the larger energy scales and gate tunability of these platforms allow for further exploration, such as through ARPES.

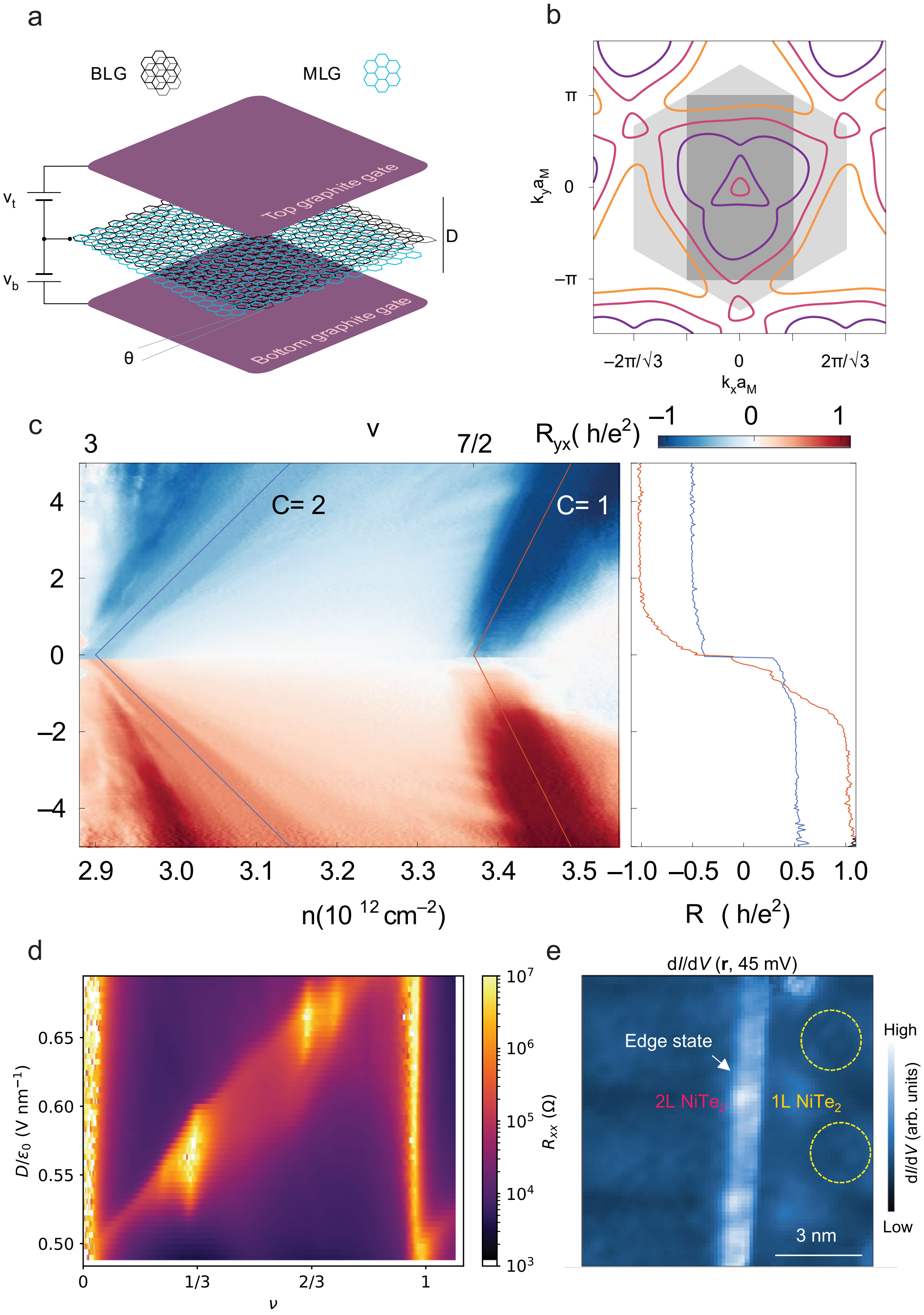

**Figure 5. Interaction-driven charge order in flat bands with nontrivial topology. (a)** Schematic of the moiré structure in a twisted monolayer-bilayer graphene (tMBG) device. **(b)** Effect of the moiré superlattice on the band structure: contour plot of the non-interacting Fermi surfaces of the C = 2 band at the graphene K valley. The purple, red, and yellow contours represent 1/4, 1/2, and 3/4 band filling, respectively. **(c)** Evolution of Hall resistance demonstrating CDW-induced tunable Chern numbers (C = 2 and C = 1) at moiré band fillings ν = 3 and 7/2 in zero magnetic field. The inset shows the corresponding quantized Hall resistance values. Panels **(a)-(c)** are reprinted with permission from H. Polshyn et al., "Topological charge density waves at half-integer filling of a moiré superlattice," Nat. Phys. 18, 42 (2022), Springer Nature. [48] **(d)** Transport signatures of charge order in a topological flat band. Longitudinal resistance map showing prominent resistive states at filling factors ν = 1/3 and 2/3, attributed to charge-density-wave states coexisting with fractional Chern insulator states. Reproduced from S. H. Aronson, T. Han, Z. Lu, Y. Yao, J. P. Butler, K. Watanabe, T. Taniguchi, L. Ju, and R. C. Ashoori, "Displacement Field-Controlled Fractional Chern Insulators and Charge Density Waves in a Graphene/hBN Moiré Superlattice," Phys. Rev. X 15, 031026 (2025), https://doi.org/10.1103/75gl-jzl6; licensed under a Creative Commons Attribution (CC BY) license (https://creativecommons.org/licenses/by/4.0/). [130] **(e)** Spatial dI/dV map at 45 mV highlighting the formation of an edge state along the boundary between monolayer and bilayer $NiTe_2$. Reprinted with permission from H. Guo et al., "Synthesis of incommensurate moiré structures with short-range-ordered charge density modulation," Nat. Commun. 16, 11327 (2025), Springer Nature. [49]

## 5. Axion-like electromagnetic response in Weyl CDW systems

The previous two sections explored topology within the context of the CDW phase and the reconstructed band structure. A third approach situates topology in the low-energy electromagnetic response, where it remains relevant even when the single-particle spectrum is fully gapped. This discussion focuses on Weyl semimetals, where low-energy quasiparticles behave as massless chiral fermions that touch at isolated nodes in momentum space, each carrying a chiral charge of $\chi = \pm 1$ [131]. When these fermions are integrated out, the electromagnetic action gains an additional term proportional to $\theta(\mathbf{r}, t)\ \mathbf{E}\cdot\mathbf{B}$, where $\mathbf{E}$ and $\mathbf{B}$ represent the electric and magnetic fields inside the material and $\theta$ is fixed by the separation of the nodes [132,133]. The terminology comes from particle physics; the axion is a hypothetical pseudoscalar field introduced to address the strong-CP problem and defined by this specific coupling to electromagnetism. This correspondence in solid-state physics applies to the effective action rather than to individual particles.

A CDW provides the dynamics that make this term significant. When the ordering wavevector Q aligns with the distance between two nodes of opposite chirality, the density wave couples states of opposite chirality, opening a gap at the nodes (Fig. 6) [132,134]. The phase of the order parameter then enters the action in place of the constant, $\theta(\mathbf{r}, t) = 2\mathbf{Q}\cdot\mathbf{r} - 2\varphi(\mathbf{r}, t)$, so that the CDW phase is promoted to a dynamical axion field. The phason — the collective mode in which the CDW slides rigidly relative to the lattice without change in amplitude [11] — then carries an anomalous contribution to the current when **E** and **B** are collinear. One would expect two consequent experimental signatures: the anomalous contribution should appear only above the depinning threshold, where the CDW slides, and it should be locked to the parallel alignment of the two fields.

Additionally, the same theory suggests that screw and edge dislocations of the density wave can form axion strings that bind gapless chiral modes [132]. This connection directly relates this class of phenomena to the defect topology discussed in Section 3.

This physics has been pursued in $Ta_2Se_8I$. Its room-temperature phase is calculated to host 24 pairs of Weyl nodes [100] and has been classified as a type-III Weyl semimetal, in which electron and hole pockets touch at a multi-Weyl point [52]; an incommensurate CDW develops below $T_{CDW} \approx 263$ K. Reports indicate that in the sliding regime of this CDW, when an electric field is applied above a specific threshold at 80 K, a collinear magnetic field contributes significantly to the magnetoconductance. This contribution exhibits a $\cos^2\varphi$ dependence on the angle φ between the electric and magnetic fields and is interpreted as the anomalous axionic contribution of the chiral anomaly to the phason current [51].

Additionally, pressure can be used to tune the properties of the same compound: it suppresses the CDW gap, and superconductivity emerges above 17 GPa. However, the Weyl semimetal phase does not survive at this pressure, and the onset of superconductivity is accompanied by the amorphization of the iodine sublattice [36]. In $(TaSe_4)_3I$, the $n = 3$ member of the same $(TaSe_4)_nI$ chain family, superconductivity below 2.5 K coexisting with weak ferromagnetism has been reported at ambient pressure [135]. Two further routes extend the category beyond a single material. Applying strain to a spatially uniform CDW in a time-reversal-invariant Weyl semimetal is predicted to generate a dynamic piezomagnetic effect through a valley axion field, without the need for phase fluctuations [106]. A CDW forming on a one-dimensional band of Landau levels underlies the proposed mechanism of the three-dimensional quantum Hall effect [99]. Field-induced chiral-symmetry-breaking density waves are also expected to occur in Weyl and multi-Weyl semimetals at weak coupling [134,136].

Establishing this class involves more than a mere coincidence of Weyl nodes and charge order in a single material. There are three essential conditions that must be met: The ordering wavevector must connect nodes of opposite chirality in the measured band structure, not just in theoretical calculations; the density-wave gap must be verified independently through spectroscopy; and the anomalous response must scale with the CDW order parameter under a tuning parameter that suppresses it, while excluding factors such as heating and inhomogeneous current paths through time-resolved and thermal analysis.

Currently, none of these conditions have been conclusively met for $Ta_2Se_8I$. Photoemission experiments show that bands crossing the Fermi level above the $T_{CDW}$ maintain their dispersion below this temperature, indicating no gap opening at the Weyl points [93]; tunneling spectroscopy at CDW edge dislocations resolves the bulk gap but no in-gap states [92]; and the collinear-field magnetoconductance has not been reproduced in subsequent measurements [90,91]. These points are discussed in more detail in Section 7. The theoretical framework for this category is well defined and distinct from the frameworks of Sections 3 and 4, but it currently relies on a single compound, with decisive measurements still pending.

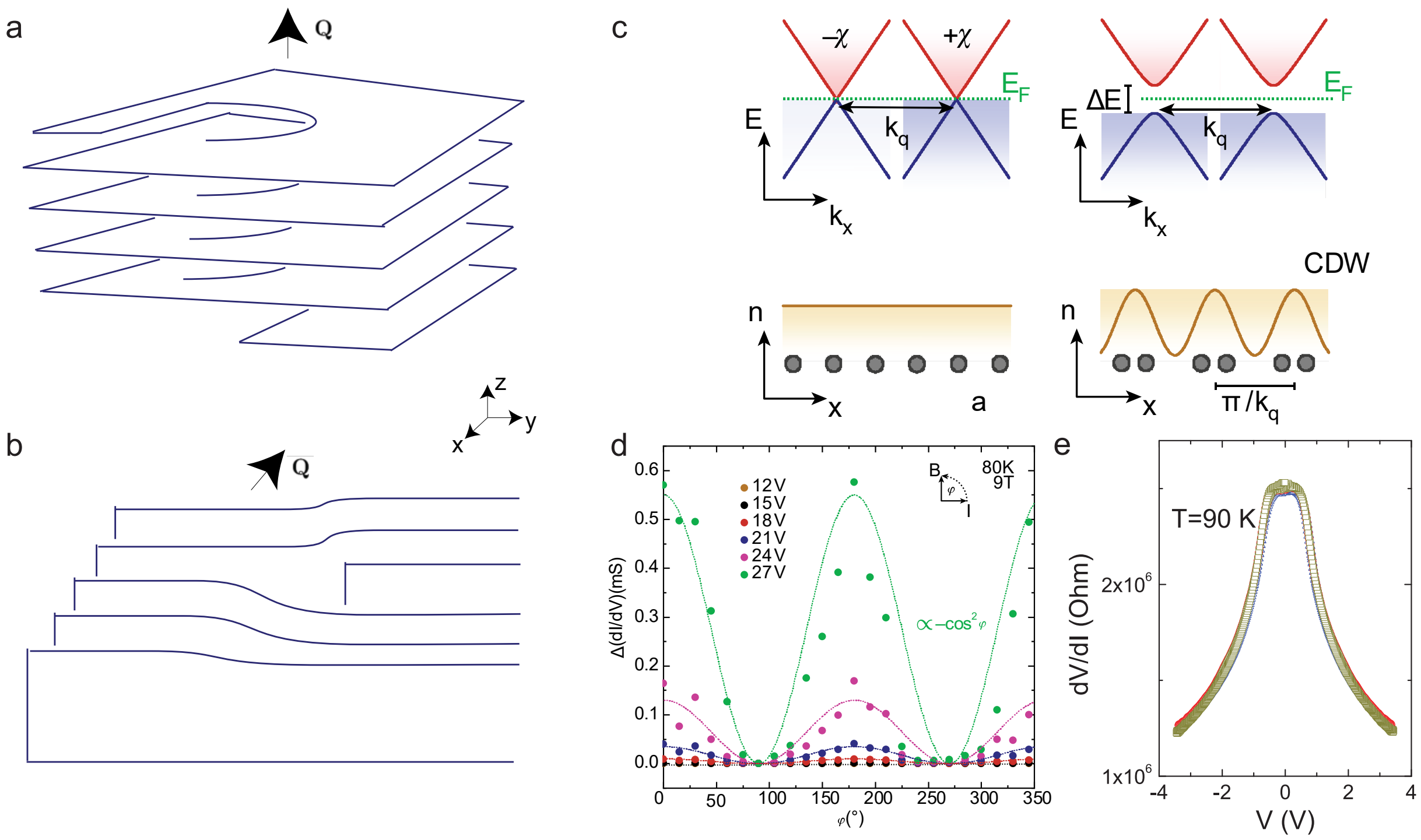

**Figure 6. Axion-like electromagnetic response in Weyl CDW systems and the contested transport evidence in $Ta_2Se_8I$.** (a)(b) Topological defects of the density wave as axion strings: (a) a screw dislocation, in which the Burgers vector is parallel to the string, and (b) an edge dislocation, in which it is perpendicular. The delineated sheets mark the maxima of the CDW and **Q** denotes the ordering wavevector; the string lies along z in both cases and binds gapless chiral modes, linking this class directly to the defect topology of Section 3. Reprinted figure with permission from Z. Wang and S.-C. Zhang, Phys. Rev. B 87, 161107(R) (2013), https://doi.org/10.1103/PhysRevB.87.161107. Copyright (2013) by the American Physical Society. (c) Mechanism of the dynamical axion field. In the undistorted phase (left) Weyl nodes of opposite chirality ∓χ are separated by $k_q$ at $E_F$ and the carrier density n(x) is uniform over the lattice period a. When the ordering wavevector matches this separation (right), the CDW couples states of opposite chirality and opens a gap ΔE, and the charge density is modulated with period $\pi/k_q$; the phase of the order parameter then enters the effective action in place of the constant θ. (d) Angular dependence of the anomalous contribution to the differential conductance, Δ(dI/dV), measured at 9 T and 80 K as a function of the angle φ between **B** and the current, for bias voltages from 12 to 27 V. Dotted lines are fits to $\cos^2\varphi$. The modulation is resolved only at the highest

biases, deep in the sliding regime. Panels (c)(d) are reprinted with permission from J. Gooth et al., “Axionic charge-density wave in the Weyl semimetal $(TaSe_4)_2I$,” Nature 575, 315 (2019), Springer Nature. (e) Differential current–voltage characteristic dV/dI(V) of $Ta_2Se_8I$ at 90 K, measured across the depinning threshold at B = 0 (dark yellow), B = 7 T parallel to the current (red), and B = 7 T perpendicular to the CDW sliding direction (blue), on crystals prepared to reproduce the geometry, contact method and measurement configuration of (d). The three curves superimpose. Reproduced from A. A. Sinchenko, R. Ballou, J. E. Lorenzo, Th. Grenet, and P. Monceau, Appl. Phys. Lett. 120, 063102 (2022), with the permission of AIP Publishing.

## 6. Nonequilibrium control and device-era

This section examines the nonequilibrium and device-level control of CDW order. It assesses whether a topological object has been identified for each phenomenon or whether the result indicates advanced CDW control with unresolved topological implications. The foundations of collective CDW transport were established more than four decades ago in transition-metal trichalcogenides. The first experimental evidence of electric-field-driven nonlinear conductivity was reported in $NbSe_3$ [137]. Further research confirmed the domain dynamics and extended to other systems such as $TaS_3$ and blue bronze $K_{0.3}MoO_3$, which were comprehensively synthesized in the seminal review by Grüner [11]. Nevertheless, such collective phenomena remained uncontrolled, limited to fundamental physics and niche device concepts.

The nonequilibrium control of the CDW phase opens a new domain of transport in low-dimensional materials. The phase of the order parameter, treated as a dynamic field, couples directly to electric fields, enabling the modification of the CDW through depinning, sliding, domain rearrangement, and hysteretic switching [61,69,71,138,139]. While nonlinear transport in CDW materials was established decades ago in quasi-1D systems [137], recent advances in exfoliation and gating of two-dimensional CDW materials have transformed the landscape, allowing direct electrostatic control of the Fermi level and, with it, the CDW ground state itself [61]. Recent work has also revealed giant gate responses in quasi-one-dimensional CDW systems, highlighting direct electric-field coupling to the CDW electron–lattice condensate [140].

1T-$TaS_2$ is among the most extensively studied charge-density-wave compounds, thanks to its rich phase diagram. Topology enters the physics of 1T-$TaS_2$ not through its band structure but through the CDW order parameter itself, the nearly-commensurate (NC) phase is threaded by $\mathbb{Z}_3$ vortices and discommensurations that organize into a honeycomb domain-wall network with emergent metallic states [141–143]. Although these topological discommensurations have not yet been used in devices, the substantial control demonstrated over the CDW phases of 1T-$TaS_2$ provides a promising foundation for their future manipulation.

1T-$TaS_2$ remains one of the most prominent platforms for controlled CDW phase switching as well. Transitions between NC and incommensurate (IC) CDW states can be driven by source-drain voltage via Joule heating [144,145], while electrostatic gating provides a more efficient and thermally decoupled control of the CDW hysteresis window (Fig. 7) [61]. In contrast, collective sliding and depinning are generally less clear-cut and less prevalent in two-dimensional CDW materials compared to classical quasi-1D systems. This is partly because higher-dimensional CDW states often involve a richer order-parameter structure and complex domain-wall networks, which complicate collective transport [146].

In 1T-$TaS_2$, the Star-of-David clusters that form in the commensurate phase create a particularly rigid pinning environment. Meanwhile, smaller commensurate domains embedded in the NC phase remain addressable by gate fields [61]. Crucially, electric-field-induced phase modulation contrasts with Joule-heating-driven switching, as the hysteresis window area remains stationary. This indicates a direct field effect on the C-CDW domain structure rather than a thermal pathway, allowing CDW switching for non-volatile memory applications [61]. Furthermore, radio-frequency excitation can modulate the NC–IC transition in 1T-$TaS_2$ by annealing frustrated CDW domain configurations, reducing domain-wall density, and reorganizing the discommensuration network. This process, in turn, alters the hysteretic current–voltage response [147]. While these control mechanisms are not inherently topological, their significance lies in the potential to manipulate the topological discommensuration network in future devices.

Beyond 1T-$TaS_2$, the CDW superconductor 2H-$NbSe_2$ has emerged as a versatile testbed for nonreciprocal transport [148]. 2H-$NbSe_2$ offers a distinct perspective: its incommensuration originates not from traditional domain walls but from the topological interplay between two coexisting, energetically competing CDW configurations [34]. Beyond its CDW state, few-layer $NbSe_2$ exhibits a pronounced supercurrent diode effect driven by valley-Zeeman spin-orbit coupling under an applied magnetic field [44]. Moreover, monolayer $NbSe_2$ supports in-plane strains large enough to produce significant band-structure modifications and strain-induced electronic phases that coexist with CDW order [149]. These latter effects provide additional control channels that could ultimately couple to the nontrivial CDW texture. Together, these responses make $NbSe_2$ a model platform for engineering coupled superconducting and charge-ordered responses within a single van der Waals material.

Optical control presents a faster method for manipulating CDWs [150–152]. Femtosecond laser pulses can drive 1T-$TaS_2$ into a metastable hidden CDW state by rearranging its domain-wall network. In this context, domain-wall vertices are characterized by their Burgers vector: topologically trivial vertices have a zero Burgers vector, while nonzero Burgers vectors indicate topologically nontrivial CDW dislocations. This technique lets switching times be set by the laser-pulse duration rather than by thermal relaxation [153].

Within the Mott insulating phase of 1T-$TaS_2$, femtosecond pulses can also influence mirror domain walls, thereby creating conduction channels along them by locally suppressing both the amplitude and the phase of the CDW [154]. This capability suggests the potential to pattern domain walls to create custom-optically functional conductive networks. Similar ultrafast engineering of topological defects and light-induced semimetallic states has been observed in 1T-$TiSe_2$ [87,155].

Vanadium chalcogenides further display this optical controllability: in 1T-$VSe_2$, femtosecond pulses transiently alter the CDW phase [156], while in $VTe_2$, where CDW order is directly linked to topological surface states [85], optical control enables the light-driven manipulation of topological band characteristics on ultrafast timescales.

Topological semimetals, including Weyl and Dirac semimetals, exhibit bulk electronic transport linked to nontrivial band topology and can support large current densities. However, their conventional transport mechanisms do not depend on the collective dynamics of an ordered condensate. In contrast, topological CDW materials provide a fundamentally different transport paradigm. In these materials, the collective nature of CDW transport serves as a control mechanism through processes such as sliding, depinning, or phase modulation. These capabilities have no equivalent in conventional topological semimetal devices.

Although true dissipationless Fröhlich transport has not yet been achieved likely due to impurity pinning, the coexistence of a sliding CDW with CDW-locked boundary states, as observed in quasi-1D materials such as $Ta_2Se_8I$ [35,100], suggests new device architectures in which CDW dynamics and the topological band structure can be modulated independently. Investigating how these two degrees of freedom interact and compete remains an exciting area for future exploration.

Looking ahead, van der Waals heterostructures present a promising avenue to couple controllable CDW order with other quantum states. This has already been demonstrated by the proximity-induced CDW order in graphene/1T-$TaS_2$ and the coexistence of superconducting and CDW layers in 6R-$TaS_2$ [157,158]. Additionally, Kagome FeGe, which exhibits the coexistence of charge order, magnetism, and a CDW-related edge state (Section 4), suggests a similar potential in magnetic topological materials.

The kagome metals $AV_3Sb_5$ exemplify the device potential arising from intertwined chiral CDW order and nontrivial topology [159]. The electrical control of quantum phase transitions has been successfully demonstrated in thin flakes, while in thicker flakes, Fermi-level shifting across the CDW gap, which achieved through proton intercalation and carrier-density modulation, induces an extrinsic-to-intrinsic anomalous Hall transition [42]. The disorder-driven superconductor-to-insulator transition [42] within this family, along with the recently demonstrated electrically switchable chiral transport [160], establishes $AV_3Sb_5$ as a promising platform for quantum devices that integrate collective phase control with chirality as a degree of freedom.

The chiral CDW state, also observed in 1T-$TiSe_2$ and 2H-$TaS_2$, may act as a precursor to odd-parity Cooper pairing and unconventional superconductivity [89,161]. Furthermore, the in-plane chirality of bulk 1T-$TaS_2$, which is controllable through shear stress and the insulating–nonconducting transition [162], suggests that field- and pulse-driven manipulation of chiral domains could enable high-speed logic gates that combine chiral degrees of freedom with nonlinear optics.

Recognizing these effects as fundamental device primitives necessitates clean signals, which places noise performance at the forefront of CDW device physics. Low-frequency noise spectroscopy has emerged as an exceptionally sensitive tool for probing CDW dynamics, effectively resolving depinning thresholds, sliding-induced fluctuations, and phase-transition anomalies within low-dimensional nanostructures. This approach directly enhances our understanding of collective and domain-scale kinetics [68,78,163].

Recent studies on $Ta_2Se_8I$ have revealed that the CDW phase of this quasi-1D topological material demonstrates remarkably low low-frequency noise during Fröhlich sliding. Notably, the normalized noise levels decrease inversely with current, contrasting with the constant noise behavior typically observed in conventional conductors [164]. Beyond its fundamental significance, this anomalously quiet sliding regime bears practical implications for quantum technologies; low-frequency noise in materials adjacent to superconducting qubits significantly contributes to flux noise and decoherence [165,166].

Thus, materials that inherently exhibit low noise during collective transport, alongside topological protection and superconductivity, may present an attractive framework for next-generation qubit architectures and cryogenic interconnects. Kagome metals such as $AV_3Sb_5$ emerge as promising candidates in this context, and mapping noise signatures to collective modes within the topological CDW family is a crucial step toward developing CDW-based quantum devices.

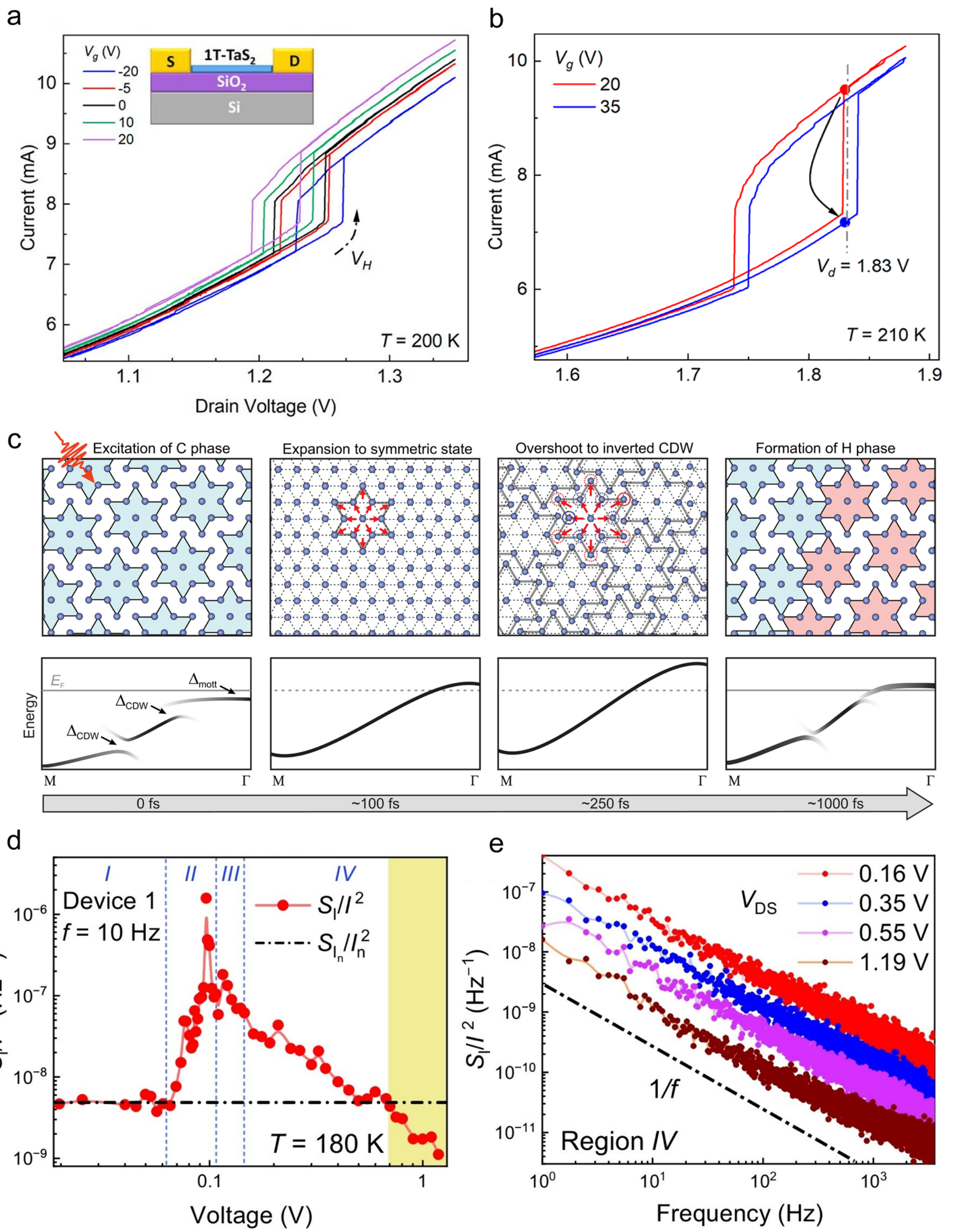

**Figure 7. Device manifestations, non-equilibrium control and noise spectroscopy of CDW phases. (a)** [69] Electrostatic gating of 1T-$TaS_2$ devices, where the threshold voltage $V_H$ for switching is tunable by gate bias. [61] **(b)** Current as a function of source–drain voltage at two fixed gate biases, showing the transition from the IC-CDW to the NC-CDW phase. Panels (a)(b) used with permission of American Chemical Society, from M. Taheri et al., "Electrical Gating of the Charge-Density-Wave Phases in Two-Dimensional h-BN/1T-$TaS_2$ Devices," ACS Nano 16, 18968 (2022); permission conveyed through Copyright Clearance Center, Inc. **(c)** Photoinduced

formation of the hidden state through collective CDW excitation, accompanied by star-of-David reconfiguration and Ta 5d band reconstruction. Reproduced from J. Maklar et al., Science Advances, DOI: 10.1126/sciadv.adi4661 (2023), AAAS. **(d)** Normalized current noise, ($S_I/I^2$), in $Ta_2Se_8I$ nanowires as a function of bias voltage, showing noise suppression in the high-bias sliding-CDW regime below the noise level of the individual charge carriers. **(e)** Frequency-dependent noise spectral density in the highlighted bias region of **(d)**, indicating transport dominated by collective CDW sliding. Panels **(d)(e)** are reproduced from S. Ghosh, N. Sesing, Z. Ebrahim Nataj, T. Salguero, S. Rumyantsev, R. K. Lake, and A. A. Balandin, "A quieter state of charge and ultra-low-noise of the collective current in quasi-1D charge-density-wave nanowires," Nat. Commun. 17, 116 (2026), https://doi.org/10.1038/s41467-025-67567-x; licensed under a Creative Commons Attribution (CC BY) license (https://creativecommons.org/licenses/by/4.0/).

## 7. Recent developments and open questions in the field

The rapid expansion of research on topological CDWs has raised new open questions. The ensuing discussions revolve around the criteria for what constitutes evidence, which observables are most decisive, and how to differentiate various notions of topology in both experimental and theoretical contexts. The first type of issue pertains to definitions. The term "topological CDW" is currently applied to at least three distinct scenarios: phase-field defects in real space, the reconstruction of electronic bands with boundary modes, and the low-energy electromagnetic response of an axionic nature.

These categories are experimentally distinguishable. Discommensurations and domain networks can be directly imaged using scanning tunneling microscopy and correlated with diffraction wavevectors, as demonstrated in studies of 2H-$NbSe_2$ [4,29,30,32,72,167]. Boundary modes tied to CDW order require spectroscopic resolution at edges or step boundaries, as demonstrated in $Ta_2Se_8I$ [35]. The axion-like response necessitates a Weyl band structure and transport signatures that trace the CDW phase, as investigated in $Ta_2Se_8I$ [36,51]. Blurring these categories can often weaken interpretation.

The second topic of debate pertains to the axionic response observed in Weyl CDW systems. In $Ta_2Se_8I$, magnetotransport measurements conducted within the CDW state were interpreted as evidence for axion electrodynamics arising from the coupling between Weyl nodes [36,51,93]. However, subsequent studies have questioned whether all observed transport features require a topological explanation, particularly given the well-established nonlinear dynamics of sliding CDWs [90–92].

This ongoing debate centers on several key issues: whether the CDW wavevector connects Weyl nodes in the actual band structure, whether the density-wave gap has been independently verified, and whether transport anomalies scale consistently with CDW phase dynamics rather than extrinsic current pathways. A resolution to this question requires a combination of band-structure calculations, spectroscopic gap measurements, and transport studies conducted under carefully controlled bias and temperature conditions.

The third topic of debate involves the symmetry characteristics of charge order in kagome metals such as $KV_3Sb_5$. Scanning tunneling microscopy has established a 2×2 superlattice and unconventional chiral charge order [40]. Subsequent experiments reported signatures interpreted as time-reversal symmetry breaking [37,38,168–170], while other measurements emphasized alternative explanations [171–173].

The central issue is whether the ordered state involves solely charge modulation or also incorporates bond currents or orbital textures that alter the Berry curvature and transport coefficients. Given that kagome systems host van Hove singularities and nontrivial band geometry near the Fermi level, minor differences in symmetry assignment can significantly influence the interpretation of anomalous Hall and Kerr signals. Thus, this issue concerns the minimal microscopic order parameter and its symmetry classification.

The fourth group of questions concerns boundary modes in CDW systems with topological band structures. In $Ta_2Se_8I$, in-gap spectral weight localized at step edges has been linked directly to CDW order [50]. The open question is whether such boundary modes are generic to CDWs in topological materials or whether they require specific symmetries and band-connectivity conditions. Establishing generality requires systematic studies across materials, including temperature-dependent boundary spectroscopy and comparison with surfaces that differ in termination and disorder.

A final topic of debate concerns the interpretation of transport in driven CDW systems. Threshold switching, oscillations, and magnetoresistance in thin films can reflect collective depinning, metastable phase transitions, or Joule heating. When such measurements are used to support topological interpretations, the burden of proof increases. The device literature in 1T-$TaS_2$ demonstrates that nanosecond pulsing [64,66,174], thermal modeling [64], and noise spectroscopy [69,70,175] are necessary to distinguish intrinsic phase dynamics from extrinsic effects. Applying the same level of control to candidate topological CDW materials remains an ongoing challenge.

The intensive debates currently shaping the field can potentially be resolved through focused experiments that integrate spectroscopy, transport measurements, symmetry analysis, and band-structure calculations. The results will ultimately clarify whether topological CDWs evolve into a well-defined category with predictive power or remain a loosely connected array of phenomena.

## 8. Outlook: what is next

Advancements in topological charge density waves now hinge on shifting from phenomenological approaches to predictive capabilities. The next phase of this field should be characterized by experiments and calculations that can definitively confirm or refute topological features with minimal ambiguity. A primary objective could be the boundary-resolved classification of CDW states in materials characterized by nontrivial band structures. The findings for $Ta_2Se_8I$ show that CDW order can significantly alter boundary spectra [35].

Systematic studies should extend this approach to additional compounds in which the band topology is independently established. The bulk CDW gap must be characterized spectroscopically. Boundary spectra must be measured with spatial resolution sufficient to track the CDW modulation. The temperature evolution of boundary modes must follow the CDW order parameter. Only when these conditions are satisfied can boundary states be attributed to the ordered phase rather than to structural termination.

The second target is quantitative control of phase defects. The commensurate–incommensurate framework developed by McMillan [28] provides a predictive relation between defect density and deviation of the ordering wavevector from commensurate values. Modern scanning tunneling microscopy and coherent diffraction methods allow direct imaging of domain walls [89,176], phase slips [177,178], and defect vertices [33,74]. Future research should combine controlled tuning parameters, such as strain or electrostatic gating, with real-space phase mapping. The goal is to correlate defect density with spectroscopic gap variations and with transport signatures. If defect-bound states contribute to in-gap spectral weight, their spatial distribution must match the phase texture. This approach connects defect topology to measurable electronic structure.

The third target focuses on the axionic response within Weyl CDW systems. For a density-wave gap to produce axion-like electrodynamics, three specific conditions must be satisfied. First, the CDW wave vector must connect Weyl nodes identified in band-structure calculations and, when possible, confirmed through angle-resolved photoemission. Additionally, the magnitude of the gap should be measured independently, for example, using tunneling spectroscopy.

Finally, it is essential to monitor magnetotransport signatures across the CDW transition and in response to tuning parameters, such as pressure, that suppress the ordered phase. An anomalous response can only be attributed to a topological mechanism if it scales with the presence and strength of CDW order. Experiments that integrate transport measurements with time-resolved control and thermal modeling will be crucial in ruling out alternative explanations.

The fourth target is symmetry-resolved identification of charge order in kagome metals. In $AV_3Sb_5$, scanning tunneling microscopy has revealed real-space modulation patterns [39,41,179]. The next step is to determine the minimal order parameter consistent with all symmetry probes, including transport, Kerr rotation, and nonlinear optical measurements. This requires coordinated experiments on the same crystals under identical conditions, with a comparison between bulk-sensitive and surface-sensitive techniques. A definitive symmetry classification will determine how charge order modifies Berry curvature and how it couples to superconductivity.

The fifth target is systematic nonequilibrium testing of topological claims. In CDW systems, an electrical bias can modify the phase field and drive transitions between metastable states. The device literature in $1T\text{-}TaS_2$ has shown that nanosecond pulsing [64,66,174], high-frequency oscillation measurements [175], and noise spectroscopy [69,70,175] can separate intrinsic collective dynamics from heating and filamentary conduction. Applying these protocols to candidate topological CDW materials will determine whether proposed boundary modes or axion-like responses persist under controlled drive. Transport measurements used to claim topological CDW must therefore be accompanied by time-resolved data and thermal analysis.

Each of these targets requires a correlation among phase order, spectroscopic gap structure, and response functions under controlled conditions. The essential tools are at our disposal: high-resolution tunneling spectroscopy, phase-sensitive imaging, band-structure calculations with symmetry analysis, and time-resolved transport in micro- and nanostructured devices. The critical advance lies in their combined application. When the CDW phase structure, boundary spectra, and electromagnetic response are measured within a cohesive framework, quantitative criteria can define topological CDWs.

We conclude by offering a discussion on how topology-related features may enhance the functionality of charge-density-wave devices. Prototype electronic architectures based on CDW materials, particularly $1T\text{-}TaS_2$, have already demonstrated voltage-controlled oscillators, transistor-less logic, and oscillatory computing platforms in which collective phase dynamics are used for information processing [180–183]. Neuro morphic and optimization-oriented oscillator networks that exploit the nearly commensurate–incommensurate (NC–IC) CDW transition have further illustrated how nonlinear phase synchronization can evolve toward solutions of benchmark optimization problems [183]. Such devices can operate at frequencies exceeding the GHz range while maintaining comparatively low switching energies, owing to the collective nature of the CDW phase dynamics.

An intriguing future possibility is that topological CDW materials may further improve device performance by reducing dissipation associated with collective transport. In materials where CDW order coexists with nontrivial band topology, symmetry-protected electronic states, modified scattering phase space, or topologically robust boundary channels may suppress selected backscattering pathways and alter the damping environment experienced by the collective CDW condensate.

While no universal topological protection against CDW dissipation has yet been established, recent observations of unusually low low-frequency noise during sliding transport in quasi-one-dimensional topological CDW systems such as $Ta_2Se_8I$ suggest that the interplay between topology and collective dynamics may already manifest in measurable transport advantages [164]. If topological band structure, chirality, or defect topology can indeed reduce effective damping, topological CDW materials may provide a pathway toward reduced-dissipation collective electronics, low-noise oscillatory networks, and energy-efficient information processing architectures that exploit collective quantum states rather than individual charge carriers.

## Acknowledgments

MSH acknowledges support from the U.S. National Science Foundation award #2610404 and DARPA award #D26AC50023.